\documentstyle[11pt,epsfig,aaspp4]{article}
\tighten
\begin{document}

\newcommand{\lya}{Lyman~$\alpha$}
\newcommand{\lyb}{Lyman~$\beta$}
\newcommand{\za}{$z_{\rm abs}$}
\newcommand{\ze}{$z_{\rm em}$}
\newcommand{\cmtwo}{cm$^{-2}$}
\newcommand{\nhi}{$N$(H$^0$)}

\def\ltsima{$\; \buildrel < \over \sim \;$}
\def\simlt{\lower.5ex\hbox{\ltsima}}
\def\gtsima{$\; \buildrel > \over \sim \;$}
\def\simgt{\lower.5ex\hbox{\gtsima}}
\def\arcs{$''~$}
\def\arcm{$'~$}
\vspace*{0.1cm}

\title{A NEW MEASUREMENT OF THE PRIMORDIAL ABUNDANCE OF DEUTERIUM:
TOWARDS CONVERGENCE WITH THE BARYON DENSITY FROM THE
CMB?\altaffilmark{1}}

\vspace{0.5cm}
\author{\sc Max Pettini}
\affil{Institute of Astronomy, Madingley Road, Cambridge, CB3 0HA, UK}

\author{\sc David V. Bowen}
\affil{Princeton University Observatory, Peyton Hall,
Princeton, NJ 08544-1001}

\altaffiltext{1}{Based on observations with the NASA/ESA
{\it Hubble Space Telescope}, obtained at the
Space Telescope Science Institute which is operated by the
Association of Universities for Research in Astronomy, Inc.,
under NASA contract NAS~5-26555.}

\begin{abstract}

From the analysis of the near-UV spectrum of the QSO
2206$-$199, obtained with a long series of exposures with STIS
on the {\it HST}, we deduce a value 
${\rm D/H} = {\rm (}1.65 \pm 0.35{\rm )} \times 10^{-5}$ 
($1 \sigma$ error) for the
abundance of deuterium in the $z_{\rm abs} = 2.0762$ damped
\lya\ system (DLA) along this sight-line. The velocity
structure of this absorber is very simple and its neutral
hydrogen column density, $N$(H~I), is accurately known; the
error in D/H is mostly due to the limited signal-to-noise ratio
of the spectrum. Since this is also one of the most metal-poor
DLAs, with metal abundances $\sim 1/200$ of solar, the
correction due to astration of D is expected to be
insignificant and the value we deduce should be essentially the
primordial abundance of deuterium. When all (six) available
measurements of D/H in high redshift QSO absorbers are
considered, we find that the three DLAs---where $N$(H~I) is
measured most reliably---give consistently lower values than
the three Lyman limit systems. We point out that the weighted
mean of the DLA measurements, 
${\rm D/H} = {\rm(} 2.2 \pm 0.2{\rm)} \times 10^{-5}$, 
yields a baryon density  $\Omega_B h^2 = 0.025 \pm 0.001$ 
which is within $\sim 1 \sigma$ of the value
deduced from the analysis of the CMB angular power spectrum,
and is still consistent with the present-day D/H and models
of Galactic chemical evolution. Future observations of D~I
absorption in other DLAs are needed to establish whether our
finding reflects a real advantage of DLAs over other classes of
QSO absorbers for the measurement of D, or is just a
statistical fluctuation.

\end{abstract}


\section{Introduction}

Among the light elements created in the Big Bang, deuterium is the one 
whose abundance depends most sensitively on the baryon-to-photon ratio, $\eta$,
which can in turn be related to the cosmological density of baryons
$\Omega_B$, through the known temperature of the cosmic
microwave background (CMB).
It was realized 25 years ago (e.g. Adams 1976) that QSO absorption line
systems at high redshift would provide the most direct avenue
to a precise determination of the primordial abundance of deuterium,
avoiding the need to account for its subsequent destruction through
successive cycles of star formation in galaxies (astration). However,
we have had to wait until the advent of 8-10\,m telescopes,
equipped with efficient high resolution spectrographs,
to realize those early expectations. 

The reason is simple.
The isotope shift between the D and H components of the Lyman  
series amounts to only $-82$\,km~s$^{-1}$; the vast majority of QSO
absorption systems span or exceed this velocity range so that features
due to D (less abundant than H by a factor of a few times $10^4$)
are normally blended with other, generally stronger, 
neutral hydrogen absorption components within the same complex. 
In order to find the 
rare absorbers with the simplest velocity structure it is necessary
to reach down the QSO luminosity function to magnitudes
which are only accessible with the light gathering power of the 
largest telescopes.
Thus, nearly ten years since the commissioning of the Keck
high resolution echelle spectrograph (HIRES; Vogt 1992)
and five years since the first detection of 
D~I absorption in a QSO spectrum (Tytler, Fan, \& Burles 1996),
the available body of data on the primordial
abundance of D is still frustratingly small, amounting to only
five measures (Kirkman et al. 2001, also reviewed below).
In contrast, determinations of the He/H ratio---far more accessible
than D/H but much less sensitive
to $\Omega_B$---number in the many tens 
(see Pagel 2000 for a recent review).

Until recently, only Lyman limit systems 
(LLS)---absorbers with neutral hydrogen column densities
$N$(H~I) $\geq 3 \times 10^{17}$\,cm$^{-2}$---had 
been targeted in searches for interstellar deuterium 
at high redshift. The rarer damped \lya\ systems
(DLAs), with $N$(H~I) $\geq 2 \times 10^{20}$\,cm$^{-2}$,
potentially offer significant advantages over 
LLS. First and foremost, the neutral hydrogen
column density can be measured more precisely
than is normally the case for LLS. The combination
of the width of the saturated core and 
damping wings of the \lya\ absorption line
usually constrain $N$(H~I) to within 10\% or better,
even in moderate resolution spectra (e.g. Pettini et al. 1997).
Second, a knowledge of the degree of metal enrichment of the
QSO absorber is required to relate the measured D/H ratio to
the primordial value. In DLAs the metallicity of the gas
can be determined straightforwardly and with good
precision, without recourse to the 
uncertain, and often large, ionization correction 
required for LLS (e.g. Viegas 1995).
Thirdly, the probability of blending with `interloper'
H~I clouds is much reduced for DLAs, given the steep
slope of the $N$(H~I) distribution (P\'{e}roux et al. 2001), and 
D absorption should be detectable in many higher order
Lyman lines, rather than in just the first few transitions
in the series.

In principle, we know of no reason why the velocity structure
of DLAs should be systematically more complex than that of LLS---one 
could in fact argue that the opposite is more likely to be the case,
if DLAs are formed in the inner regions 
of protogalactic disks while LLS arise in more extended
halo regions (e.g. Steidel 1993). 
Furthermore, there is mounting evidence,
both observational (e.g. Pettini et al. 1999 and references therein)
and theoretical (e.g. Mo, Mao, \& White 1998; Jimenez, Bowen, \& 
Matteucci 1999) that DLAs may trace preferentially
galaxies with low rates of star formation, where 
astration of D is also likely to have been low.
In any case, we believe that DLAs are no less suitable than LLS
in the search for high redshift D~I absorption; they are just
rarer and this explains why they have not figured as prominently
up to now.

Among the well studied DLAs, the $z_{\rm abs} = 2.0762$
system in the bright (V = 17.3) $z_{\rm em} = 2.559$ QSO
Q2206$-$199 is a prime candidate for showing resolved
D~I absorption. High resolution spectroscopy
obtained initially with the University College London echelle
spectrograph at the Anglo-Australian telescope
(Pettini \& Hunstead 1990) and later with 
HIRES on Keck (Prochaska \& Wolfe 1997)
showed that in this DLA all the metal lines 
from neutral gas apparently 
consist of a single, narrow component
with a Doppler parameter $b = 5.7 $\,km~s$^{-1}$.
This is also one of the lowest metallicity
DLAs known, with different elements 
being underabundant relative to solar
by factors between $170$ (Si) and $420$ (Fe).
Given the minimal amount of chemical enrichment
experienced by the interstellar medium of this
absorber, we could reasonably expect its deuterium 
abundance to be very close to the primordial value.

At the high column densities of DLAs, the D and H components
are saturated and blended together in the strongest lines of the 
Lyman series. In order to resolve the two isotopes,
it is necessary to access high order transitions;
at $z_{\rm abs} = 2.0762$ all the lines of diagnostic value 
fall below the atmospheric cut-off at
3100~\AA\ and are not accessible from the ground.
For this reason in December 1996
we were awarded 31 orbits of the {\it Hubble Space Telescope}
to record the near-UV spectrum of Q2206$-$199 
with STIS (the Space Telescope Imaging Spectrograph).
For a variety of reasons the full set of observations 
was only completed four years later, in September 2000.
In this paper we present the resulting co-added spectrum 
from which we obtain a new
measurement of the deuterium abundance.
In the mean time, D~I absorption has been detected
in another DLA, the  $z_{\rm abs} = 3.0249$
system in Q0347$-$393, by D'Odorico, Dessauges-Zavadsky, \& Molaro
(2001) who used the Ultraviolet-Visual Echelle Spectrograph (UVES)
on the Very Large Telescope (VLT) of the European Southern Observatory.
Thus, the total number of D/H measures in high redshift QSO
absorbers now stands at six; we briefly comment on the 
dispersion of these values and on what it may imply for an improved
determination of the primordial abundance of deuterium.\\

\section{Observations}
The STIS spectra were recorded with the near-UV MAMA detector 
and G230M grating through the $52 \times 0.2$~arcsec long slit
with 31 orbits spread over seven visits between 
November 13, 1998 and September 16, 2000.
The total integration time was 82\,490~s. The grating was set 
at a central wavelength $\lambda_{\rm c} = 2818$\,\AA.
With this configuration, the useful wavelength range
recorded by the near-UV MAMA detector is 2773--2862\,\AA;
in the rest frame of the $z_{\rm abs} = 2.0762$ DLA this corresponds
to the interval 901.4--930.4\,\AA\ which includes all Lyman series lines from
Ly7 to the Lyman limit. The resolution is 0.19\,\AA\ (20\,km~s$^{-1}$ 
FWHM) sampled with 2.1 pixels.

The 31 STIS exposures were processed 
with the standard CALSTIS pipeline software.
We extracted one-dimensional spectra from
the fluxed and wavelength calibrated frames with the IRAF
routine {\texttt apall} and then added them together
after rebinning to a common wavelength scale.
We found that the pipeline processing underestimates
the scattered light in these low level exposures;
consequently, we applied
an empirical background correction
of 5\% of the continuum level in order to bring the cores of the 
saturated absorption lines to zero.
Figure 1 shows the final spectrum, background subtracted, normalized to the QSO continuum,
and reduced to the rest frame of the $z_{\rm abs} = 2.0762$ DLA.
At the low light levels of the present observations the noise is dominated
by the dark count of the NUV MAMA (which turned out to be 
significantly higher than pre-flight expectations and more than one hundred times
higher than that of the FUV MAMA) and is roughly constant
along the spectrum, irrespectively of the QSO signal.
Its magnitude can be best appreciated from the 
fluctuations about the zero level in the cores of the strong
saturated absorption lines near 927.5 and 923\,\AA\ and below 
the effective Lyman limit at 914\,\AA.
In the continuum near Ly9 (see below) the signal-to-noise ratio
is ${\rm S/N} \simeq 10$.\\

\section{Deuterium in the $z_{\rm abs} = 2.0762$ DLA}
As can be appreciated from Figure 1, Q2206$-$199
exhibits a crowded spectrum at UV wavelengths. 
The QSO flux is absorbed by
a multitude of features due to the \lya\ forest, 
the $z_{\rm abs} = 2.0762$ DLA,
and a second DLA intercepted in this direction, 
at $z_{\rm abs} = 1.9205$, with much stronger
and wider interstellar absorption lines (Pettini \& Hunstead 1990;
Prochaska \& Wolfe 1997).
For the present purposes we focus on the high order
Lyman series of the $z_{\rm abs} = 2.0762$ DLA
from Ly7 at 926.2257\,\AA\ to Ly13 at 916.429\,\AA;
transitions to energy levels with $n \geq 14$
are so close in wavelength that they
become blended with one another eventually producing an effective
Lyman break near 914.5\,\AA.
Figure 2 shows on an expanded scale
six of the seven Lyman lines considered here.
Three of them, Ly10, Ly11, and Ly13,
are blended with other absorption features, as is Ly8
(shown in Figure 1). The remaining three lines, however,
appear to be relatively free of contamination and D~I absorption
is clearly present in Ly7, Ly9, and Ly12. These are the
three spectral features on which our measurement
of $N$(D~I) is based.

All the six Lyman lines in Figure 2 were fitted with
Voigt profiles using the VPFIT package.\footnote{VPFIT
is available at http://www.ast.cam.ac.uk/\~\,rfc/vpfit.html}
VPFIT returns the most likely values of 
redshift $z$, Doppler width $b$ (km~s$^{-1}$),
and column density $N$ (cm$^{-2}$)
by minimizing the difference 
between observed and computed profiles
after convolution with the appropriate instrumental
point spread function (PSF). We used the STIS PSF
for the G230M grating
supplied by the Space Telescope Science Institute;
close to the line core the function is well
approximated by a Gaussian with FWHM = 20\,km~s$^{-1}$
but its wings are significantly broader.
In our fitting procedure we fixed the column density
of neutral hydrogen to $N$(H~I)\,$= 2.73 \times 10^{20}$~cm$^{-2}$
deduced from the profile of the damped \lya\ line
(see \S4 below), and the number of absorption components to one.
All the metal absorption lines from H~I gas in this
DLA are best reproduced by
a single absorption component at $z_{\rm abs} = 2.07623$ 
with Doppler parameter $b = 5.7$\,km~s$^{-1}$
(Prochaska \& Wolfe 1997; Pettini et al. in preparation).
Since the same value of $b$ apparently applies to
atoms spanning a range of atomic weight from 14 (N) to
56 (Fe), this velocity dispersion presumably reflects primarily turbulent, 
rather than thermal motions.
Accordingly, we fixed $b_{\rm turb} = 5.0$\,km~s$^{-1}$
and let VPFIT solve for the thermal component, $b_{\rm th}$, to the 
H~I and D~I Lyman lines; the results are collected in Table 1.
The theoretical absorption profiles corresponding to the 
best solution returned by VPFIT are shown as continuous lines 
in Figure 2.

We deduce log$N$(D~I)\,$= 15.65 \pm 0.1$ ($1 \sigma$).
The error is dominated by the modest signal-to-noise ratio
of the STIS spectrum, rather than by systematic uncertainties
in the continuum level or the details of the fitting procedure.
We explored at length the effects of adopting different
values of $b_{\rm turb}$, varying the continuum level and $N$(H~I) by 
$\pm 1 \sigma$, assuming a two-component model for
the absorption lines\footnote{As appropriate to the high
ionization gas traced by the Si~IV and C~IV lines in this
DLA}, and in all cases VPFIT 
converged to values of log$N$(D~I) which were well within the 
$\pm 0.1$~dex uncertainty due to the noise in the spectrum.
Figure 3 illustrates how the theoretical profiles obtained
by lowering and increasing the value of $N$(D~I)
by $2 \sigma$ compare with the observed spectral lines
in the two most sensitive cases, Ly9 and Ly12.
The fits shown include absorption from
adjacent components which are not D~I;
excluding these components resulted in values of $N$(D~I)
which are within the error quoted (although the match to
the observed profiles was naturally worse).

We note in passing that $b = 14.6$\,km~s$^{-1}$ (Table 1)
corresponds to $b_{\rm th} = 13.7$\,km~s$^{-1}$
for our assumed $b_{\rm turb} = 5.0$\,km~s$^{-1}$
(since $b_{\rm th} \gg b_{\rm turb}$, 
very similar values of $b_{\rm th}$ are obtained 
for all possible values of $b_{\rm turb}$, from 
0 to 5.7\,km~s$^{-1}$).
The implied temperature of the H~I gas,
$T = m\,b^2/2k$ where $m$ is the atomic mass
and $k$ is Boltzmann constant, is $T = 11\,300$\,K.
This value is essentially the same as that deduced in 
an analogous way
by O'Meara et al. (2001) in their determination
of the abundance of D in the $z_{\rm abs} = 2.536$
sub-DLA ($N{\rm (H~I)} = 2.65 \times 10^{19}$\,cm$^{-2}$)
toward the QSO HS~0105+1619.
Temperatures of $\sim 10^4$\,K may appear rather high
for H~I gas. However, it is now well established that
high-$z$ DLAs have far higher spin temperatures
(deduced by comparing 21\,cm and \lya\ absorption)
than H~I clouds in the Milky Way. The recent 
comprehensive compilation of values of $T_s$ by Kanekar \& Chengalur (2001)
includes eight DLAs with {\it lower limits}
ranging from $T_s > 800$\,K to $T_s > 4700$\,K.
Both the DLA considered here and the one studied by
O'Meara et al. (2001) have metallicities
$Z \simlt 1/100 Z_{\odot}$. In this regime the low
cooling rate from metals may well result in interstellar
H~I temperatures as high as $10^4$\,K. Alternatively,
the relatively large velocity dispersions deduced by ourselves
and O'Meara et al. (2001)
may be indicative of the presence of additional
absorption components, but in this case it would be a coincidence
that both DLAs yield similar values of $T$.
We stress again, however, that the value of $N$(D~I)
derived above does not depend on the precise
details of the velocity structure of the absorber
because the lines most sensitive to $N$(D~I),
Ly9 and Ly12, are both on the linear part of the 
curve of growth.\\

\section{The Column Density of H~I and the Abundance of Deuterium}
Figure 4 shows the normalised 
spectrum of Q2206$-$199 in the region near 3740\,\AA,
which encompasses the damped \lya\ line at $z_{\rm abs} = 2.07623$. 
This spectrum was obtained by Pettini et al. (in preparation)
as part of a program to study the abundances of N and O in DLAs.
The profile of the DLA is best fitted with a column density
$N{\rm (H~I)} = 2.73 \times 10^{20}$\,cm$^{-2}$
(top panel of Figure 4).
VPFIT returned a formal error $\sigma_N$ of less than 1\%.
Even though the strong \lya\ line is relatively unblended,
we consider this estimate to be over-optimistic.
The main source of error here is probably the continuum
placement {\it within} the damped line
(from $\sim 1200$ to $\sim 1235$\,\AA---see Figure 4);
various trials with different interpolations across this 
wavelength interval
showed that $\pm 5 \times 10^{18}$\,cm$^{-2}$ is a more
realistic assessment of the $1 \sigma$ error. 
In any case, the error on $N$(H~I) is clearly 
much smaller than the error on $N$(D~I) and it is
the latter that dominates our measure of D/H.
For comparison, the lower resolution and S/N spectrum
of Q2206$-$199 obtained by Pettini et al. (1994),
with a different detector and telescope, yielded
$N{\rm (H~I)} = {\rm (}2.7 \pm 0.4{\rm )} \times 10^{20}$\,cm$^{-2}$,
in excellent agreement with the value deduced here.
Evidently, the column density of neutral gas in this DLA 
is known with a satisfactory degree of confidence for our
present purpose.

From the above measurements we deduce a value of the abundance
of deuterium ${\rm D/H} = {\rm (}1.65 \pm 0.35{\rm )} \times 10^{-5}$;
this is the lowest estimate of this ratio obtained up to now.\\

\section{Discussion}
At the time when the observations presented here
were first proposed, it was still being debated 
whether the primordial
abundance of deuterium was a few times $10^{-5}$ or 
one order of magnitude higher (Tytler, Fan, \& Burles 1996; 
Songaila et al. 1995).
The controversy is now largely resolved;
with several more low values of D/H
reported in the last few years, it seems
increasingly likely that cases where D/H
is apparently greater than $10^{-4}$
are in reality due to contamination
by H~I at velocities similar to the isotope shift---always
a possibility in the \lya\ forest (e.g. Kirkman et al. 2001).
The finding here of a sixth system with low D/H
adds further weight to this conclusion. 

In Table 2 we have collected published determinations
of D/H in QSO absorption line systems at high $z$;
the data are mostly from a similar compilation by O'Meara et al. 
(2001---their Table 5) augmented by the recent determinations
by D'Odorico et al. (2001) and ourselves.
The results are plotted in Figure 5. Although the
set of measurements is still very limited,
we comment on two points of potential interest.

All six absorbers are low metallicity systems, with abundances
(as measured by Si) less than 1/30 of solar ([Si/H]\,$< -1.5$). 
We do {\it not}
believe that Figure $5a$ shows a trend of decreasing D/H with
increasing [Si/H]. The main reason is that in a closed-box
model of chemical evolution a metallicity $Z =  -1.5$
is reached when, to a rough approximation, only 3\%
of the gas has been processed through stars---and only 3\% of the
deuterium has been destroyed. Thus an increase in metal abundances
from $-2.5$ to $-1.5$ should result in an imperceptible
decrease in the D/H ratio through astration. 
An additional consideration to be
aware of is that not all measurements of [Si/H] in Figure $5a$
are equally reliable. While the metallicities of the three DLAs
(squares) are reasonably secure, the estimates in the three LLS
(triangles) rely on much larger ionization corrections to
account for unobserved ion stages. Thus one should be wary of
any apparent correlation of D/H with $Z$ when results from
LLS and DLAs are mixed together.

For these reasons it has been assumed 
until now that in all the cases listed in Table 2 we are
measuring the primordial abundance of deuterium and that
the dispersion between values found in different QSO absorbers
is just due to observational error. Thus, Tytler et al. (2000)
and O'Meara et al. (2001) proposed averaging the available
determinations to obtain (in the latter, more recent work) 
a most likely value
${\rm D/H} = {\rm (}3.0 \pm 0.4{\rm )} \times 10^{-5}$. 

On the other hand, it is interesting that all three
DLAs give consistently lower values of D/H than
the three LLS (Figure $5b$). 
In general DLAs allow a more accurate measure
of the column density of H~I than LLS; the shape of the damping wings
and the core of the \lya\ line normally constrain the allowed
values of $N$(H~I) to a narrow range.
This is not always the case for LLS where $N$(H~I) 
relies more sensitively on the accuracy of the extrapolation
of the QSO continuum near the Lyman limit and on the details
of the velocity structure of the H~I gas giving rise to saturated 
Lyman series lines. To illustrate this point, we note that
in PKS~1937$-$1009 the initial measure 
$N{\rm (H~I)} = {\rm (}8.7 \pm 1.2{\rm )} \times 10^{17}$\,cm$^{-2}$
by Tytler, Fan, \& Burles (1996)
was later revised by Burles \& Tytler (1997) down to 
$N{\rm (H~I)} = {\rm (}7.2 \pm 0.3 {\rm )} \times 10^{17}$\,cm$^{-2}$,
while a different analysis (Songaila, Wampler, \& Cowie 1997)
proposed $N{\rm (H~I)} < 5 \times 10^{17}$\,cm$^{-2}$.
In Q1009$+$299 there is the additional problem of partial
contamination of the D~I absorption by unrelated H~I
(Burles \& Tytler 1998); this is usually less of a concern
in the higher column density damped systems.

We do not intend to be critical of these excellent analyses
which have taken great care to account for all the relevant
factors and arrived at the best estimates allowed by the data.
Here we simply speculate, on the basis of the 
apparent trend in Figure $5b$, that D/H may have been {\it
over}estimated in the two LLS where deuterium has been detected
so far, possibly through an {\it under}estimate of the column
density of neutral hydrogen, and that the most reliable measure
of the primordial D/H may in fact be provided by the three damped
\lya\ systems, where $N$(H~I) is most secure. Clearly this
conjecture will be tested by future observations of additional
DLAs; nevertheless, it is worthwhile considering 
briefly its implications, should it turn out to be correct.

From the three DLAs in Table 2 we obtain a weighted mean:
\begin{equation}
{\rm D/H} = {\rm(} 2.2 \pm 0.2 {\rm)} \times 10^{-5}
\end{equation}
where the weights used are inversely proportional
to the square of the error on each measurement
and $0.2 \times 10^{-5}$ is the $1 \sigma$ error on the weighted mean.
Using the analytic expression by Burles, Nollett, \& Turner
(2001), the primordial abundance of deuterium in eq. (1)
implies a baryon-to-photon ratio:
\begin{equation}
\eta = {\rm (}6.8 \pm 0.4{\rm )} \times 10^{-10}
\end{equation}
which in turn yields a baryon density
\begin{equation}
\Omega_B h^2 = 0.025 \pm 0.001
\end{equation}
using the scaling of $\Omega_B h^2$ with $\eta$
given by Burles et al. (2001).

We note that a primordial abundance of deuterium as low as
${\rm D/H} = {\rm (}2.2 \pm 0.2{\rm )} \times 10^{-5}$ can
still be reconciled with the present-day value 
${\rm (D/H)}_{\rm ISM} = {\rm (} 1.6 \pm 0.1 {\rm )} \times 10^{-5}$ 
in the Galactic interstellar medium (Tytler et al. 2000 and references
therein), particularly when the unrealistic (for the Milky Way
at least) assumption of closed-box chemical evolution is
relaxed to include outflows and infall of unprocessed gas (e.g.
Edmunds 1994; Tosi et al. 1998). 
It is also possible that ${\rm (D/H)}_{\rm ISM}$ is less than
$1 \times 10^{-5}$, given two recent lower determinations
(Vidal-Madjar et al. 1998; Jenkins et al. 1999).
Turning to other light elements produced in the Big Bang,
the value of $\eta$ in eq. (2) implies a primordial\/ $^4$He
abundance $Y_P = 0.249$ which is still within the range
of current estimates (Pagel 2000), but a high value of the primordial
Li/H\,$ = 6.0 \times 10^{-10}$ which exacerbates the well-known 
lithium problem (Ryan et al. 2000; Burles et al. 2001).

In the last few months there has been considerable discussion
of a possible conflict between the values of $\Omega_B$ deduced
from the abundances of the light elements interpreted within
the framework of Big-Bang nucleosynthesis on the one hand, and
from the analysis of the angular power spectrum of the 
CMB on the other. Recent measurements
of the latter have been analyzed to give a best fitting
\begin{equation}
\Omega_B h^2 \simeq  0.032^{+0.005}_{-0.004}
\end{equation}
where the error is again $1 \sigma$ (Jaffe et al. 2001).
The $2 \sigma$ difference from the baryon density implied 
by existing measurements of D/H as summarized by Tytler
et al. (2000) and O'Meara et al. (2001) has prompted
(not for the first time) speculations that `new physics'---or
at least previously unaccounted for processes (e.g. Naselsky
et al. 2001)---may be required to reconcile these two 
determinations. On the other hand, if the low values
of D/H found in damped \lya\ systems are indeed 
representative of the true primordial abundance of D,
as speculated here, the apparent conflict is reduced
considerably.
The two estimates in eqs. (3) and (4) are nearly within
$1 \sigma$ of each other
and the history of astronomical measurements
teaches us that $1 \sigma$ errors almost invariably
turn out to be over-optimistic in the light
of subsequent improvements in the data.

The key question is whether the systematically low values
of D/H in the three DLAs in Figure 5
are a significant result or just a statistical accident.
This question can only be answered by more observations
of DLAs with the characteristics required to detect and measure
D~I absorption. Fortunately the prospects for finding such
valuable test cases are excellent, given the
spectacular increase in the numbers of known QSOs 
brought about by large-scale searches such as the Sloan
digital sky survey and 2dF QSO survey.
A significant investment in observing time will be 
required to follow up these QSOs with higher resolution 
spectroscopy. However, it is now clear that only 
by assembling a moderately large sample
of D/H measurements at high redshift will it be possible
to differentiate between systematic and random errors and 
eventually determine the primordial abundance of deuterium.\\

\section{Epilogue}

Since this paper was submitted for publication there have
some important developments of relevance to the discussion
above. Three new estimates of $\Omega_B$ have been announced,
all based on observations of the CMB power spectrum. 
From the analysis of the full dataset of the BOOMERANG
experiment, Netterfield et al. (2001) and de Bernardis et al.
(2001) deduced $\Omega_B h^2 = 0.021^{+0.004}_{-0.003}$
($1 \sigma$ errors).
The first results from the Degree Angular Scale Interferometer
(DASI) also give $\Omega_B h^2 = 0.022^{+0.004}_{-0.003}$
(Pryke et al. 2001). 
A new treatment of the MAXIMA-I
measurements favours a value which is higher,
$\Omega_B h^2 = 0.0325 \pm 0.007$, but still
consistent with the other two
within the errors (Stompor et al. 2001).
Thus one could reasonably argue that the `problem'
has been solved and that $\Omega_B$ is now known to 
within $\sim 10$\%,
since the weighted average of 
all five detections in Table 2, 
${\rm D/H} = {\rm (}2.6 \pm 0.2{\rm )} \times 10^{-5}$,
implies $\Omega_B h^2 = 0.022 \pm 0.001$, in excellent
agreement with the latest CMB mesurements.
It just remains to be seen how this newly found concordance 
stands the test of future measurements.

A second recent development is the re-analysis by Levshakov
et al. (2001) of the absorption
lines in the $z_{\rm abs} = 3.0249$ DLA in Q0347$-$383
previously studied by D'Odorico et al. (2001). By including in
the complex velocity structure of this DLA a narrow ($b =
3$\,km~s$^{-1}$) component whose presence is suggested by S~II
and H$_2$ absorption lines, Levshakov et al. propose that the
abundance of D should be revised upwards from (D/H) = ($2.25
\pm 0.65$) $\times 10^{-5}$ (the value derived by D'Odorico et
al. and quoted in Table 2) to (D/H) = ($3.2 \pm 0.4$) $\times
10^{-5}$. If this is indeed the case, then the evidence for a
systematic difference in the values of (D/H) between  Lyman
limit and damped \lya\ systems is made significantly
weaker. As emphasized above, only further searches for D in
absorption systems with a {\it simple} velocity structure,
such as that presented here, will clarify this point.

In any case, as far as primordial nucleosynthesis is
concerned, one of the most urgent goals now is identifying the
origin of the dispersion in the values of D/H in Table 2 and
Figure 5. The simplest explanation is that the errors of the
individual measurements have been underestimated. However, it
is worrying that a similarly wide dispersion in D/H can be
found in the local ISM, where the factor of $\sim 2$ variation
indicated by earlier work (Sonneborn et al. 2000 and
references therein) is apparently confirmed by the latest,
high precision, observations with the {\it Far Ultraviolet
Spectroscopic Explorer} (Moos et al. 2001). At both high and
low redshift, we do not have a satisfactory explanation for
differing D/H ratios in gas of similar chemical composition.
Resolving this puzzle is important for two reasons. First, it
dents our confidence in the general framework of Big-Bang
nucleosynthesis. Second, it thwarts future attempts to use the
abundance of D, together with those of elements which are
manufactured---rather than destroyed---by stars, to track the
chemical evolution of galaxies over the age of the universe.
Such attempts seem somewhat futile at present given that the
total amount of astration of D over the Hubble time is
comparable to the scatter in the determination of D/H at any
redshift.\\

We are most grateful to Bob Carswell, who kindly performed for
us the line profile fits presented in this paper, and to Bernard
Pagel for helpful comments. MP would like to acknowledge
J. Bergeron, P. Petitjean, S. Ellison, and C. Ledoux, 
his collaborators in the VLT program to study DLAs;
the data shown in Figure 4 were obtained as part of that study.

\newpage


{}

\newpage

%
%


\begin{deluxetable}{lccc}
\tablewidth{17.0cm}
\tablecaption{BEST FITTING PARAMETERS FOR H I and D I ABSORPTION AND $1 \sigma$ ERRORS}
\tablehead{
\colhead{Ion} & \colhead{$N$ (cm$^{-2}$)} & \colhead{$b$ (km~s$^{-1}$)} &
\colhead {$z$}
}
\startdata
H I  & ${\rm (}2.73 \pm 0.05{\rm )} \times 10^{20}$$^a$  & $14.6 \pm 0.2^b$ & $2.076234 \pm 0.000002$   \nl
D I  & ${\rm (}4.5 \pm 1{\rm )} \times 10^{15}$        & $10.6^{c}$       & $2.076234^{d}$  \nl
\enddata
\tablenotetext{a}{Determined from the profile of the damped \lya\ line}
\tablenotetext{b}{The Doppler parameter $b$ has both thermal and turbulent components
($b^2 = b_{\rm th}^2 + b_{\rm turb}^2$); $b_{\rm turb}$ was fixed at $5$\,km~s$^{-1}$
(see text)}
\tablenotetext{c}{Fixed to be $b$(D~I)$^{2} = b_{\rm turb}^2 + 1/2 \times b_{\rm th}$(H~I)$^{2}$}
\tablenotetext{d}{Fixed to be the same as $z$(H~I)}
\end{deluxetable}


\begin{deluxetable}{lccccc}
\tablewidth{17.0cm}
\tablecaption{SUMMARY OF D/H MEASUREMENTS AT HIGH REDSHIFT}
\tablehead{
\colhead{QSO} & \colhead{$z_{\rm abs}$} & \colhead{$N$(H~I) (cm$^{-2}$)} & 
\colhead{[Si/H]$^a$}   & \colhead{(D/H)$\pm 1 \sigma$ ($10^{-5}$)}    &\colhead{Ref.}
}
\startdata
Q0130$-$403  &  2.799 & $4.6 \times 10^{16}$ &  $-2.6$    & $< 6.8$        & O'Meara et al. (2001) \nl
Q1009$+$299  &  2.504 & $2.5 \times 10^{17}$ &  $-2.53$   & $4.0 \pm 0.65$ & O'Meara et al. (2001) \nl
PKS 1937$-$1009 & 3.572 & $7.2 \times 10^{17}$ & $-2.26$  & $3.25 \pm 0.3$ & O'Meara et al. (2001) \nl
HS~0105+1619 &  2.536 & $2.6 \times 10^{19}$ &  $-2.0^b$  & $2.5 \pm 0.25$ & O'Meara et al. (2001) \nl
Q2206$-$199  &  2.0762 & $2.73 \times 10^{20}$ & $-2.23^c$ & $1.65 \pm 0.35$ & This paper \nl
Q0347$-$383  &  3.025  & $4.3 \times 10^{20}$  & $-1.53^d$ & $2.25 \pm 0.65$ & D'Odorico et al.(2001) \nl
\enddata
\tablenotetext{a}{In the usual notation, [Si/H] = log(Si/H) $-$ log(Si/H)$_{\odot}$. We chose
Si as an indicator of the overall metallicity $Z$ because this is the element for which 
most measurements are available for the absorption systems considered here}
\tablenotetext{b}{This is actually [O/H]. The upper limit [Si/H]\,$< -1.85$ determined
by O'Meara et al. (2001) is consistent with the oxygen abundance}
\tablenotetext{c}{Prochaska \& Wolfe (1997)}
\tablenotetext{d}{Prochaska \& Wolfe (1999)}
\end{deluxetable}

%
%

%
%

\begin{figure}
\figurenum{1}
\vspace*{-3.5cm}
\hspace*{-2.50cm}
\psfig{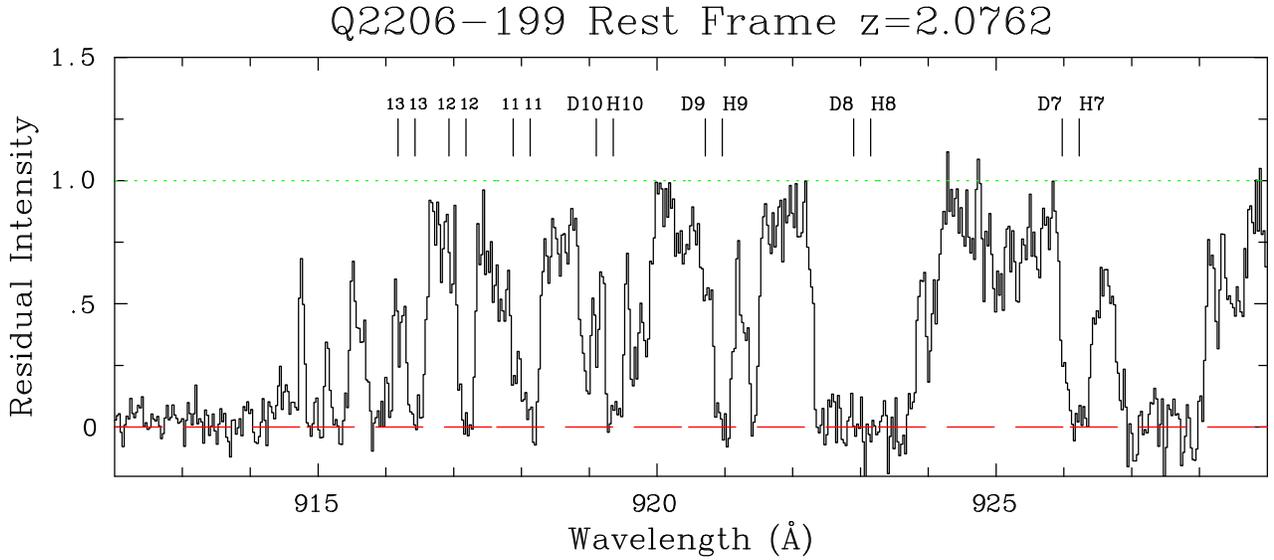}
\vspace{-12.0cm}
\figcaption{Near-UV (2805.5--2857.8\,\AA) STIS spectrum of Q2206$-$199
normalised to the QSO continuum and reduced to the rest frame of the 
$z_{\rm abs} = 2.0762$ DLA. The spectral resolution is
20\,km~s$^{-1}$ FWHM 
and ${\rm S/N} \simeq 10$; the total exposure time was 82\,490~s.
The locations of high order Lyman lines of H~I and D~I, from Ly7 to Ly13,
are indicated by vertical tick marks above the spectrum.
}
\end{figure}

%
%

\begin{figure}
\figurenum{2}
\vspace*{-0.75cm}
\hspace*{-2.0cm}
\psfig{figure=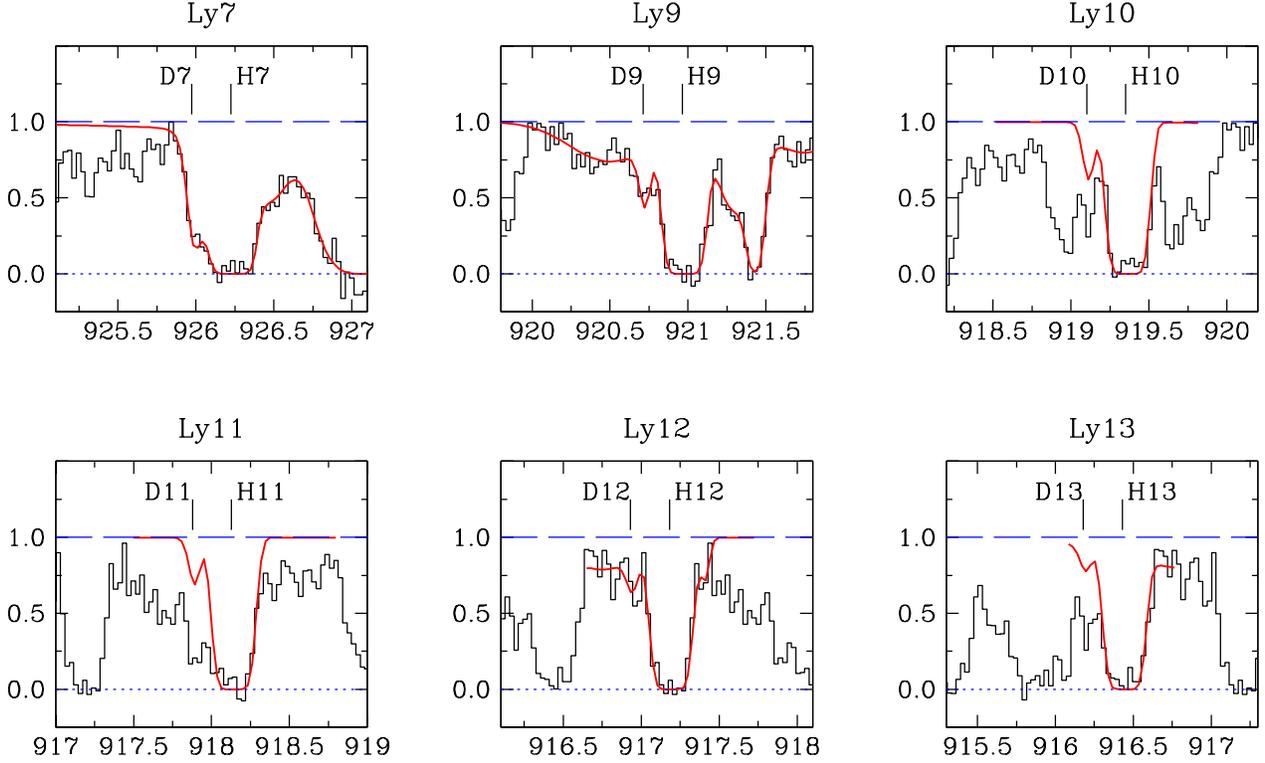,width=145mm,angle=270}
\vspace{-1.25cm}
\figcaption{
Residual intensity vs. rest wavelength (\AA) for
high order Lyman lines in the $z_{\rm abs} = 2.0762$ DLA.
The histograms are the data, while the thin continuous lines 
show the theoretical absorption profiles for H~I and D~I 
corresponding to the best fit
solution returned by VPFIT (see Table 1). 
The important lines here are Ly7, Ly9, and Ly12 which
are sufficiently free of contamination 
to allow a determination of the column density
of D~I. The fits shown for these three lines include adjacent,
unrelated, absorption features.
For Ly10, Ly11, and Ly13 we show only the contributions
of the H~I and D~I components to the complex blends  
present near their wavelengths.
Although there is absorption at the expected position of D~I
in each of these lines, the blends 
in these three cases are 
{\it dominated} by contaminating features,
as evidenced by the fact that
the optical depths do
not match the relative $f$-values of Ly10, Ly11, and Ly13.
}
\end{figure}

%
%

\begin{figure}
\figurenum{3}
\psfig{figure=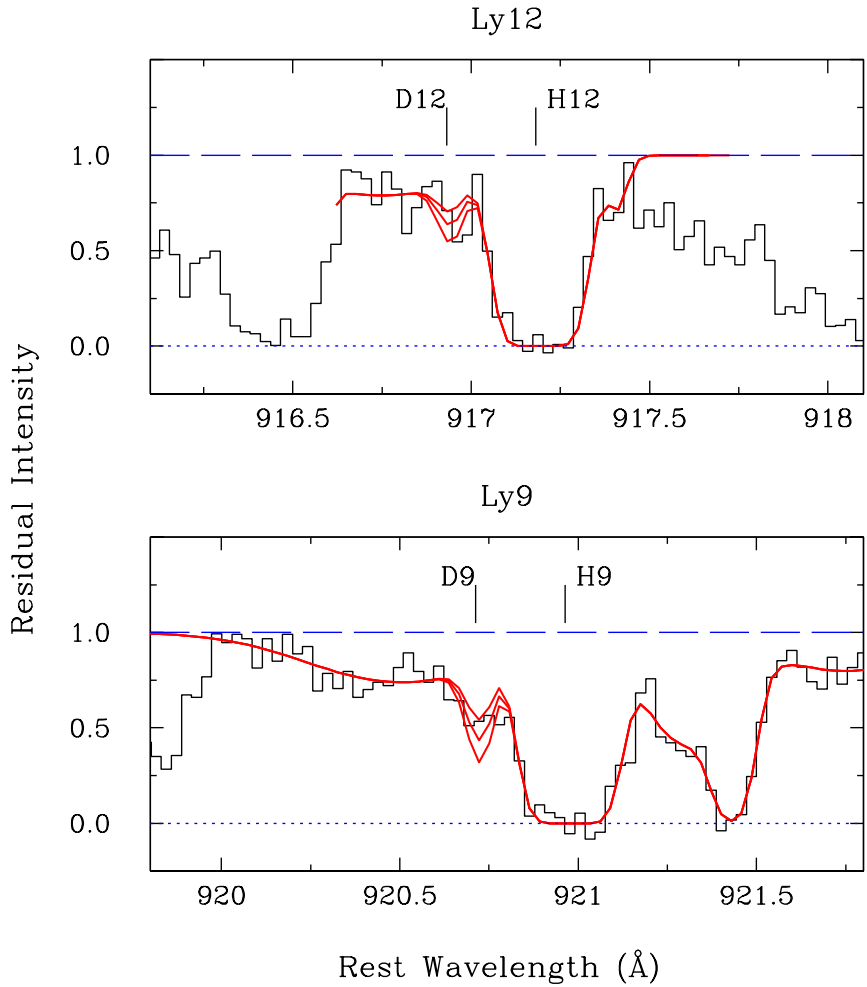,width=155mm}
\vspace{-5.0cm}
\figcaption{Comparison between observed (histogram)
and computed (continuous line) profiles for 
log$N$(D~I)$\, = 15.65 \pm 0.2$. This range
corresponds to {\it twice} the $1 \sigma$ error
on $N$(D~I) returned by VPFIT.
}
\end{figure}

%
%

\begin{figure}
\figurenum{4}
\vspace*{-0.5cm}
\psfig{figure=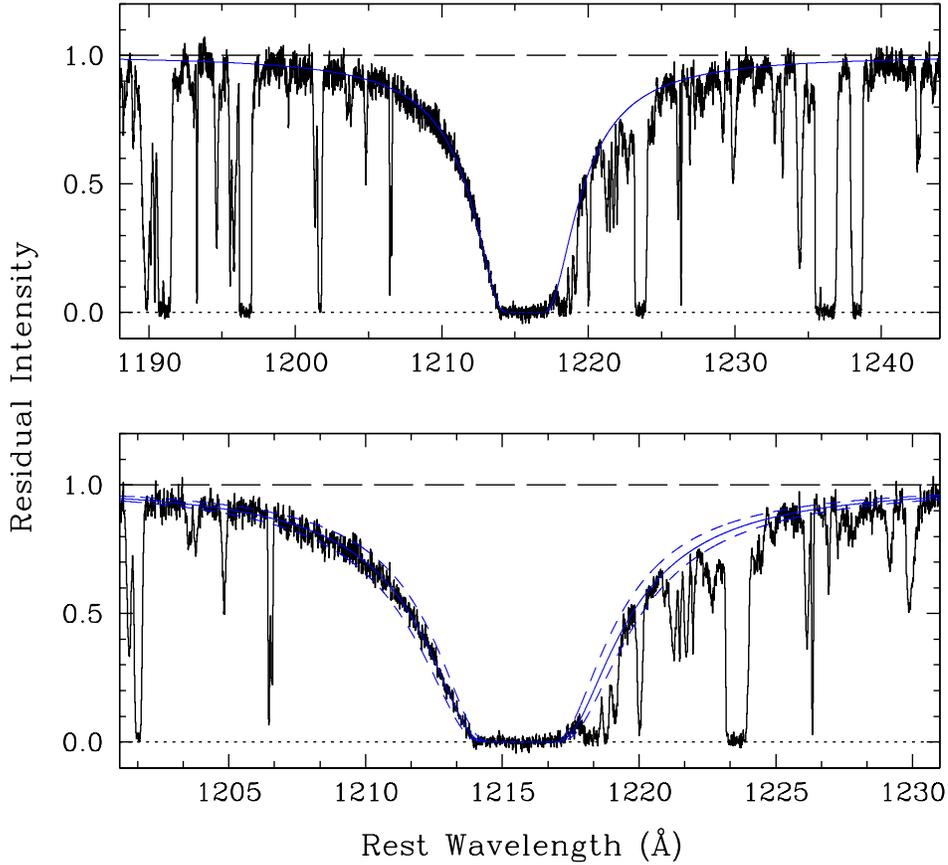,width=155mm}
\vspace{-4.5cm}
\figcaption{Damped \lya\ line in the $z_{\rm abs} = 2.0762$ DLA in Q2206$-$199
obtained with UVES on the VLT
(reproduced from Pettini et al. in preparation).
The resolution is 7\,km~s$^{-1}$~FWHM.
The fit shown in the top panel is for $N{\rm (H~I)} = 2.73 \times 10^{20}$~cm$^{-2}$.
The lower panel shows the damped line on an expanded scale together 
with three fits, corresponding to 
$N{\rm (H~I)} = {\rm (}2.73 \pm 0.5{\rm )} \times 10^{20}$~cm$^{-2}$.
The range shown corresponds to 10 times the estimated error
on $N$(H~I).
}
\end{figure}

%
%

\begin{figure}
\figurenum{5}
\vspace*{-0.5cm}
\psfig{figure=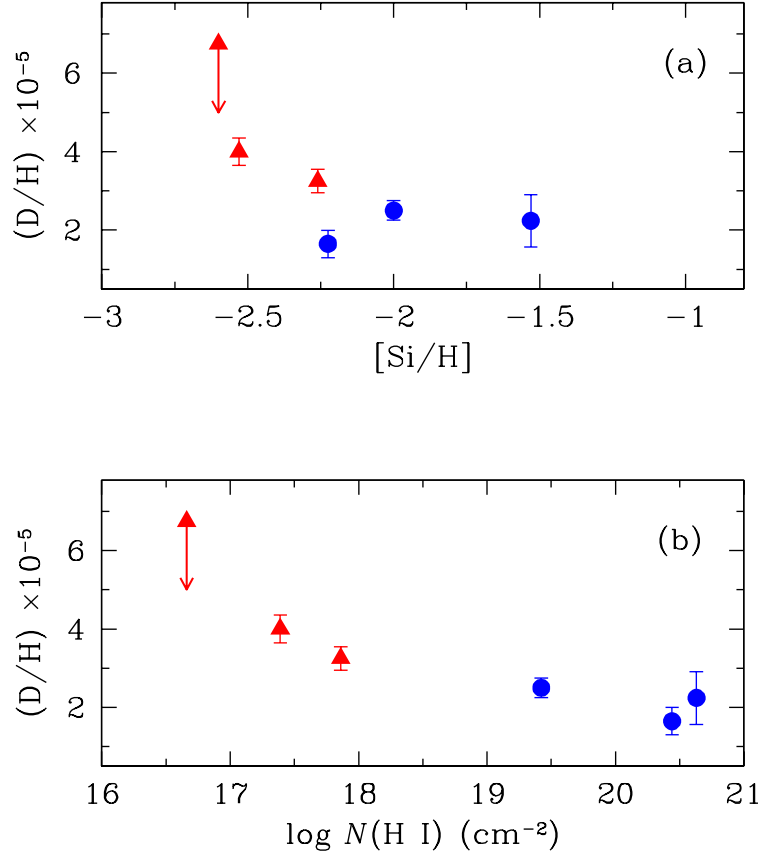,width=155mm}
\vspace{-4.5cm}
\figcaption{Available measurements of the abundance of deuterium
in QSO absorption systems at high redshift from Table 
2---{\it triangles:}\/ Lyman limit systems; {\it circles:}\/ damped
\lya\ systems.
The top panel shows D/H as a function of metallicity, as measured by the
[Si/H] ratio. In the lower panel the deuterium abundance
is plotted against the column density of neutral gas.
Damped \lya\ systems seem to yield systematically lower
values of D/H than Lyman limit systems.
}
\end{figure}

\end{document}